# Absorption mode broadband 2D MS for proteomics and metabolomics

*Maria A. van Agthoven[1,2,3\*], Marek Polák[2,4], Jan Fiala[2,4], Claude Nelcy Ounounou[1], Petr Halada[2], Michael Palasser[3], Anne Briot-Dietsch[5], Alan Kádek[2,4], Kathrin Breuker[3], Petr Novák[2,4], Carlos Afonso[1], Marc-André Delsuc[5]*

[1]Université de Rouen-Normandie, Institut CARMeN UMR 6064, IRCOF, 1 rue Tesnière, 76821 Mont St Aignan Cedex, France

[2]Institute of Microbiology of the Czech Academy of Sciences, Videnska 1083, Prague 14220, Czech Republic

[3]Center for Chemistry and Biomedicine, University of Innsbruck, Innrain 80/82, 6020 Innsbruck, Austria

[4]Faculty of Science, Charles University, Hlavova 6, Prague 12843, Czech Republic

[5]Institut de Génétique et de Biologie Moléculaire et Cellulaire, INSERM U596, CNRS UMR 7104, Université de Strasbourg, 1 rue Laurent Fries, 67404 Illkirch-Graffenstaden, France

KEYWORDS

Proteomics, metabolomics, tandem mass spectrometry, FT-ICR MS, two-dimensional mass spectrometry

ABSTRACT

Two-dimensional mass spectrometry (2D MS) is a method for tandem mass spectrometry that enables the correlation between precursor and fragment ions without the need for ion isolation. On a Fourier transform ion cyclotron resonance mass spectrometer, the phase correction functions for absorption mode data processing were found to be linear in the precursor ion dimension and quadratic in the fragment ion dimension. Absorption mode data processing on limited data sets has previously shown improvements in signal-to-noise ratio and resolving power by a factor of 2. Here, we have expanded absorption mode data processing to 2D mass spectra regardless of size and frequency range. We have applied absorption mode 2D MS to top-down analysis of variously oxidized ubiquitin proteoforms generated by fast photochemical oxidation of proteins (FPOP) and to an extract of ergot alkaloids. We show that absorption mode data processing significantly improves both the signal-to-noise ratio and the resolving power of the 2D mass spectrum compared to standard magnitude mode in terms of sequence coverage in top-down proteomics, as well as the accuracy of precursor-fragment correlation in metabolomics.

INTRODUCTION

Two-dimensional mass spectrometry (2D MS) is a method for tandem mass spectrometry that enables the correlation between precursor and fragment ions without the need for ion isolation.[1] On a Fourier transform ion cyclotron resonance mass spectrometer (FT-ICR MS), precursor ion radii in the ICR cell are modulated by using a pulse sequence developed by Pfändler *et al* with an incremental delay.[2–4] The modulation frequency is the cyclotron frequency of the ions minus an offset. Gas-free fragmentation methods in the ICR cell yield fragmentation efficiencies that are dependent on precursor ion radii.[5,6] When precursor ion radii are modulated before fragmentation, fragment ion abundances are conversely modulated with the same frequency as their precursors, allowing for their correlation. A aata set consists of transients recorded for each value of a series of regularly incremented delays. After Fourier transformation of each transient, a second Fourier transformation is performed according to incremental delay. The result is the 2D mass spectrum, which shows the fragmentation profiles of the ions from the analytes. To date, 2D MS has been performed with a wide variety of fragmentation methods compatible with high vacuum in the ICR cell: infrared multiphoton dissociation (IRMPD), electron capture dissociation (ECD), electron-induced dissociation (EID), electron-detachment dissociation (EDD), or ultraviolet photodissociation (UVPD), demonstrating its applicability to a range of analytical scenarios.[7–10]

In one-dimensional FT-ICR MS, phase correction for absorption mode improves both the signal-to-noise ratio (SNR) and the resolving power of the mass spectrum.[11,12] The phase of signals in the transient is found to be a quadratic function of cyclotron frequency.[13] The zero-order, first-order, and second-order coefficients of the phase function have been found to depend mostly on instrument parameters.[14] Multiple programs and algorithms have been developed to calculate phase correction functions and process mass spectra in absorption mode as easily as possible.[15,16]

In contrast, for 2D MS, the phase correction function was found to be quadratic in the fragment ion frequency dimension and linear in the precursor ion frequency dimension, with the coefficients in the two dimensions independent from each other.[17] Similarly to 1D MS, the first 2D mass spectrum to be processed in phase-corrected absorption mode showed improved SNR and resolving power in both dimensions.[18] Crucially, absorption mode 2D MS was used to accurately correlate precursors and their fragments for a mixture of two histone peptides that differed only by *m/z* 0.006 (acetylation vs. trimethylation), demonstrating the immense power hidden in this kind of analysis.[19] However, unlike the automated magnitude mode data processing of 2D mass spectra, which relies on batch-processing, the Spectrometry Processing Innovative Kernel (SPIKE) software (the most widely used algorithm for 2D MS) required loading the entire data set in a computer's random-access memory (RAM), thereby limiting the size of data sets that could be processed in absorption mode.[20] Until now, absorption mode was only applied to narrowband mode 2D mass spectra, with precursor ion frequency ranges under 20 kHz.[21] Most 2D mass spectra are acquired in broadband mode, with precursor ion frequency ranges over 150 kHz to simultaneously fragment the whole mass range of analytes. Therefore, to be truly useful, phase correction for absorption mode data processing requires expansion to all frequency ranges and data sizes.

One of the current promising applications of 2D MS is top-down analysis of covalently modified biomolecules. Label-free relative quantification in 2D MS has enabled the identification, location, and quantification of modified peptides.[21] Recently, 2D MS of chemically-labelled ubiquitin tracked the solvent accessibility of its lysine residues.[22] This result opens up the possibility of using 2D MS for the top-down analysis of proteins with other covalent-labelling techniques, such as Fast Photochemical Oxidation of Proteins (FPOP), in which hydroxyl radicals are generated with an excimer laser in a quench flow setup to oxidize solvent-accessible residues of proteins in a native-like environment.[23] In most applications, oxidized proteins are digested before analysis by liquid chromatography coupled with tandem mass spectrometry (LC-MS/MS).[24] Recently, FPOP has been combined with top-down analysis by direct infusion and tandem mass spectrometry on FT-ICR MS.[25–27] The top-down analysis of FPOP-oxidized ubiquitin by 2D MS generated fragmentation patterns from multiple proteoforms, which enabled phase-correction for absorption mode processing.

Here, we address the limitation of absorption mode data processing to only narrowband mode 2D MS and RAM-limited data sizes with an update to the automated data processing program in SPIKE. We showcase its performance using a broadband 2D mass spectrum of FPOP oxidized ubiquitin and we compare the results to the same data set processed in magnitude mode. To demonstrate the wide applicability of this approach, we also performed phase-corrected absorption mode processing to a 2D UVPD mass spectrum of an extract of ergot alkaloids to examine the performance improvement of phase-corrected absorption mode for 2D MS in both top-down proteomics and metabolomics.

EXPERIMENTAL METHODS

*Sample preparation*

Ubiquitin from bovine erythrocytes was purchased from Sigma-Aldrich (Saint-Louis, MO, USA) and oxidized using the FPOP setup described by Yassaghi *et al.*[25] The solution was diluted to a 1 µM final protein concentration in aqueous solution of 1% acetic acid and 50% methanol for analysis. The extract of ergot alkaloids was prepared in a 1:10 vol. solution of methanol. All solvents were LC-MS grade and obtained from Merck, Darmstadt, Germany.

*Instrumental parameters*

All 2D MS experiments were performed on a 15 T solariX FT-ICR mass spectrometer (Bruker Daltonik, Bremen, Germany) with an electrospray ion source operated in positive mode and direct infusion at a flow rate of 2 µL/min.

The pulse sequence for the 2D MS experiment is shown in Scheme 1. For ECD fragmentation, the two pulses in the encoding sequence (precursor detection and modulation) were set at 5.02 dB attenuation with 1.0 µs per excitation frequency step (frequency decrements were 625 Hz). The corresponding amplitude was estimated at 250 $V_{pp}$, with a 1.9% sweep excitation power for an amplifier with a maximum output of 446 $V_{pp}$. For UVPD fragmentation, the two pulses were set at 15.86 dB attenuation (estimated amplitude of 72 $V_{pp}$, with an 0.4% sweep excitation) after optimization for the UVPD fragmentation zone in the ICR cell. In the horizontal fragment ion dimension, the excitation pulse in the detection sequence for all 2D mass spectra was set at 3.09 dB attenuation with a 20 μs/frequency step (frequency decrements were 625 Hz). The corresponding amplitude was estimated at 312 $V_{pp}$, with a 35% sweep excitation power for an amplifier with a maximum output of 446 $V_{pp}$.

For the 2D mass spectrum of oxidized ubiquitin, ions were accumulated for 0.1 s before being transferred to the dynamically harmonized ICR cell (Paracell). The encoding delay $t_1$ was increased 4096 times with a 2 μs increment, which corresponds to a 250 kHz frequency range. No phase-cycled signal averaging was employed in the experiment. The minimum cyclotron frequency for the modulated precursor ions was 92.2 kHz for a maximum $m/z$ of 2500 during excitation, leading to a $m/z$ 673.31–2500 mass range for precursor ions. Captured ions were fragmented by ECD using the following parameters: the hollow cathode current was 1.5 A, the ECD pulse length 15 ms, the ECD lens 25 V, and the ECD bias 1.5 V.[28] The horizontal mass range was $m/z$ 207.274–2500 (corresponding to a frequency range of 1111.1–92.1 kHz). Transients were acquired over 0.236 s with 512k data points. The total duration of the experiment was 40 min.

For the 2D mass spectrum of the ergot alkaloid extract, continuous accumulation of selected ions (CASI) was used with two windows: $m/z$ 550 (Δ$m/z$ 100) and $m/z$ 810 (Δ$m/z$ 380) with 50 ms accumulation for each window to partially suppress the prominent base peak at $m/z$ 610.[29] The encoding delay $t_1$ was increased 4096 times with a 2 μs increment, which corresponds to a 250 kHz frequency range. No phase-cycled signal averaging was employed in the experiment. The minimum cyclotron frequency for the modulated precursor ions was 230.3 kHz for a maximum $m/z$ of 1000 during excitation, leading to a $m/z$ 479.52–1000 mass range for precursor ions. Captured ions were fragmented by a single 5 mJ pulse of a 193 nm excimer laser (Excistar 500, Coherent, Saxonbourg, USA), custom-coupled to the mass spectrometer through a $BaF_2$ window at the rear of the ICR cell. The horizontal mass range was *m/z* 115.15–1000 (corresponding to a frequency range of 2000.0–230.3 kHz). Transients were acquired over 0.131 s with 512k data points. The total duration of the experiment was 38 min.

**Scheme 1.** Pulse sequence of the 2D MS experiment.

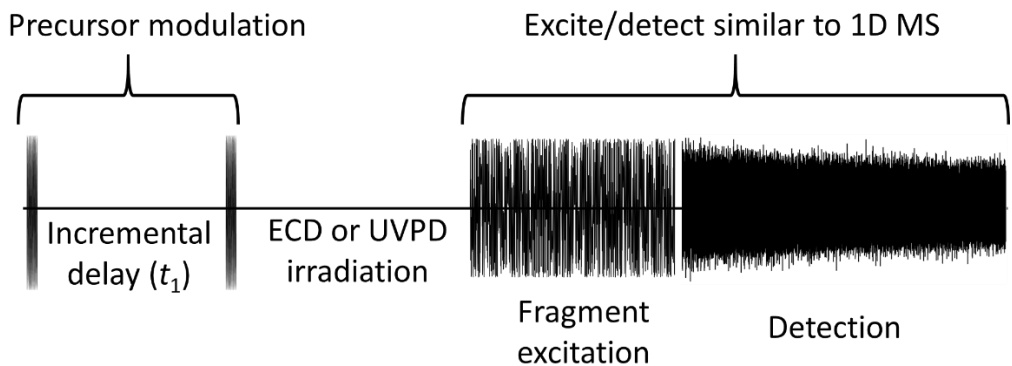

Data processing

The two-dimensional mass spectra were processed and visualized using SPIKE (available at www.github.com/spike-project, version 0.99.33, accessed on July 1, 2024) developed by the University of Strasbourg (Strasbourg, France) and CASC4DE (Illkirch-Graffenstaden, France) in the 64-bit Python 3.10 programming language on an open-source platform distributed by the Python Software Foundation (Beaverton, OR, USA).[20,30] Processed data files were saved using the HDF5 file format.

The 2D mass spectra were processed in both magnitude and absorption mode. In both cases, batch processing was used to avoid loading the entire data set in RAM. Magnitude mode processing has been described previously.[31] For phase-corrected absorption mode processing, each transient was apodized with a variant on the asymmetric apodization function proposed by Kilgour and Van Orden for absorption mode FT-ICR MS without baseline correction.[32] Then, a quadratic phase correction function was applied to the frequency-domain spectrum of each transient. The coefficients of the phase correction function were optimized on a one-dimensional mass spectrum with the same experimental conditions for the excite-detect part of the pulse sequence. In the vertical precursor ion dimension, a linear correction function was applied with coefficients that were optimized with a precursor ion scan of a fragment resulting from multiple precursor ions.[17,18] The baseline of each precursor ion scan was corrected by recalculating the initial data points with an autoregressive model discussed in the results.

Processing was performed with and without Support Selection for Noise Elimination (SANE) denoising with the same apodization and zerofilling.[30,33] The program used for data processing in magnitude mode is available in SPIKE.[20] The program used for data processing in absorption mode is available with the raw data set, alongside the configuration files for both magnitude mode and absorption mode processing of the 2D mass spectrum. In absorption mode, both processed 2D mass spectra had 8192 by 1M datapoints. In magnitude mode, the 2D mass spectrum of oxidated ubiquitin had 8192 by 1M datapoints and the 2D mass spectrum of ergot alkaloids had 8192 by 512k datapoints.

Data analysis

For each precursor ion species recorded in the 2D mass spectrum of oxidized ubiquitin, five fragment ion scans were added up to cover the precursor isotopic distribution and obtain isotopic distributions for all fragment ions as has been described in a previous study.[22] The resulting one-dimensional fragment ion patterns were peak-picked in SPIKE. Frequency-to-mass conversion was quadratic in the horizontal fragment ion dimension.[34] Peak assignments were performed using the Free Analysis Software for Top-down Mass Spectrometry (FAST-MS) developed by the University of Innsbruck (Innsbruck, Austria) in the 64-bit Python 3.7 programming language.[35,36] FAST-MS generated theoretical c/z and a•/y fragment lists for ubiquitin variably modified with 1-2 oxidations located on the following residues: phenylalanine, histidine, isoleucine, lysine, leucine, methionine, proline, arginine, valine, threonine, and tyrosine.[26,37]

For the 2D mass spectrum of the ergot alkaloid extract; peak-picking and centroiding were performed with a two-dimensional peak-picker in SPIKE and quadratic frequency-to-mass conversion was performed on the fragment ion scan of ergocryptine (m/z 576.3) and applied in both dimensions.

RESULTS AND DISCUSSION

Top-down analysis of a mixture of proteoforms provides a germane test case for phase corrected absorption mode broadband 2D MS. To estimate the coefficients of the quadratic phase correction function in the horizontal dimension, a one-dimensional MS/MS spectrum with the same excite/detect parameters as the 2D MS pulse sequence (see scheme 1) can be used, as shown in previous studies.[17–19] Because in top-down analysis multiple charge states of very similar analytes are fragmented, a common fragment originating from enough precursors with a wide *m/z* range to estimate the coefficients of the linear phase correction function in the vertical dimension is easily found.[38] The coefficients used for absorption mode in 2D MS are listed in the configuration file for data processing. In applications with a wider diversity of analytes, where fragment ions present fewer precursor ion peaks in the 2D mass spectrum, an alternative strategy to estimate the coefficients of the linear phase correction function may be to sum multiple precursor ion scans prior to phase correction. Since the phase of fragment ion signals evolves the same way according to $t_1$ (see Scheme 1) regardless of the fragment, we can hypothesize that this strategy enables the accurate estimation of the phase correction coefficients.[17]

In Figure 1, we show some of the fragmentation patterns of the 9+ charge state of ubiquitin after FPOP in the 2D mass spectrum after magnitude mode processing (Fig. 1a) and absorption mode processing (Fig. 1b). The fragmentation patterns of charge states ranging from 10+ to 6+ can be seen in Figure S1 in the Supporting Information. Both data processing methods show that linear phase correction is accurate in broadband mode over a frequency range of 250 kHz, which had only been adequately shown in narrowband mode with a 10 kHz frequency range.[18,19] 2D MS experiments in previous studies using phase-corrected absorption mode were also limited in size because the entire file had to be loaded in the RAM for data processing. The program for absorption mode processing in this study was modified so that it only requires loading in RAM of a single row or column of the data set at a time, which essentially removes this limit.

Figs. 1a and 1b are both contour plots. The lowest contour level is set slightly above noise level. Both figures show fragment ion peaks for the unmodified ubiquitin as well as ubiquitin bearing either one or two oxidations. Both the resolving power of the peaks and their signal-to-noise ratios are significantly higher in absorption mode than in magnitude mode, as predicted. In the horizontal dimension, the resolving power at $m/z$ 390 was measured to increase from 79,000 in magnitude mode to 188,000 in absorption mode. The resolving power at $m/z$ 857 in the vertical dimension was measured to be 1,200 in magnitude mode and 2,900 in absorption mode. For the M+3 isotopic peak of $z_{73}^{7+}$ (mostly containing 2 $^{13}$C isotopes) at $m/z$ 1168.8477 from the $[M+9H]^{9+}$ precursor at $m/z$ 952.69, the SNR was measured to be 24 horizontally and 16 vertically in magnitude mode. For the same peak in absorption mode, the SNR was 80 horizontally and 43 vertically. In each case, the resolving power and SNR were at least doubled. Because the data is hypercomplex, the theoretical gain in resolving power and SNR is 2, and the slight difference between the experimental and theoretical values explained by the fact that magnitude mode and absorption mode processing involved different apodization functions.[1,39]

The observed improvement in resolving power and SNR leads to more fragment ions being detected and assigned with improved confidence (e.g. $a_{31}^{3+}$, $a_{21}^{2+}$ for [M+9H]$^{9+}$, and $(a_{51}+O)^{5+}$, $(c_{51}+O)^{5+}$, $(z_{62}+O)^{6+}$ for the [M+9H+O]$^{9+}$ precursor in Fig. 1). For comparison, a recent publication by Rahman *et al.* sought to improve the SNR of a 2D EDD mass spectrum of oligonucleotides by accumulating 8 scans for each value of $t_1$ instead of 1 scan, with magnitude mode data processing.[10] This method multiplied the experiment time and sample consumption by a factor of 8 and only improved the SNR by a factor of 1.3-3.0 without any improvement in resolving power.

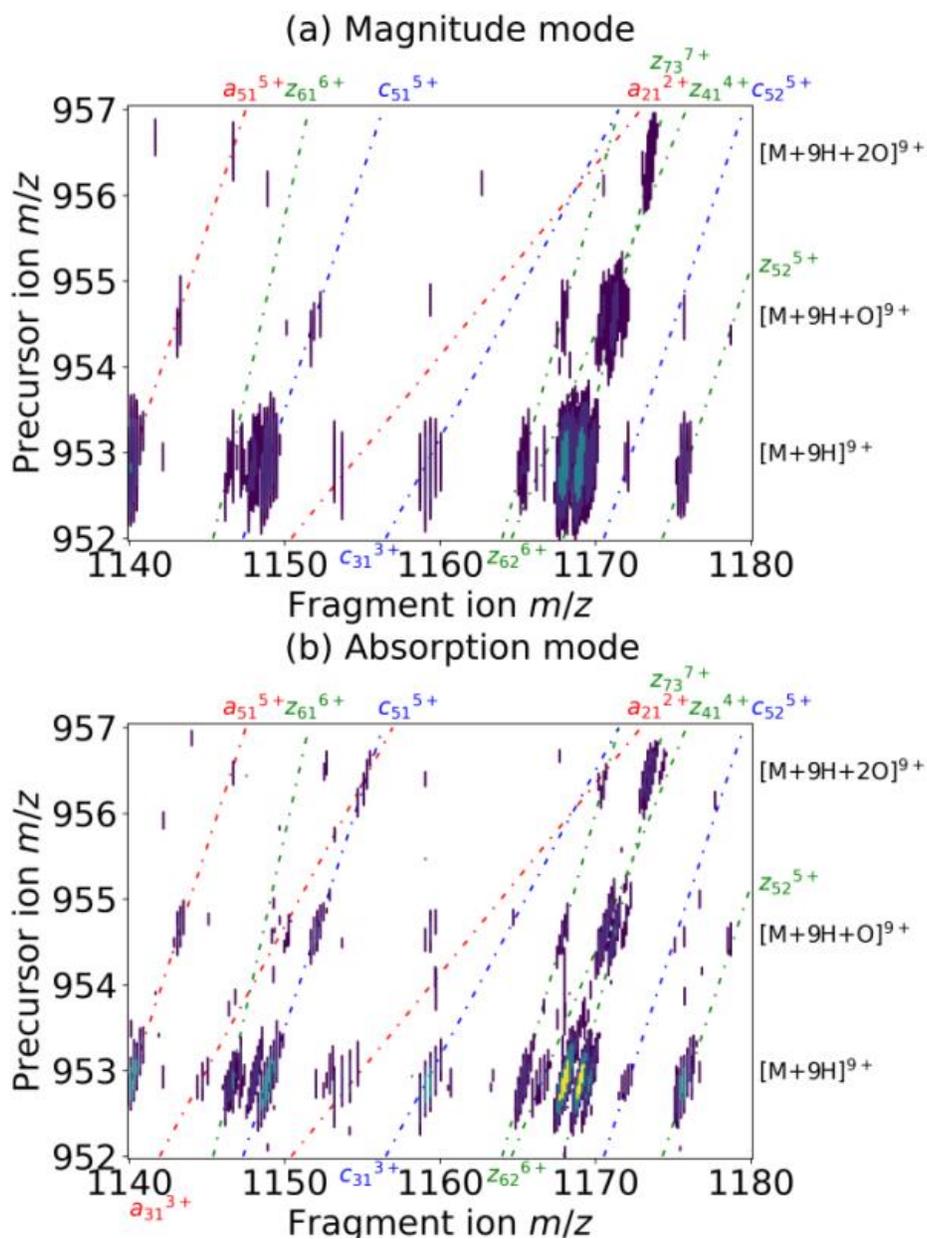

**Figure 1.** 2D ECD mass spectrum of oxidized ubiquitin after (a) magnitude mode processing and (b) absorption mode processing. Dissociation lines are shown in dotted lines (red: *a*-type fragmentations, blue: *c*-type fragmentations, green: *z*-type fragmentations). For clarity, only a zoomed-in area of the 2D mass spectrum is presented. Areas of the 2D mass spectra with the fragmentation patterns of multiple charge states are shown in the Supporting Information (Fig. S1).

One issue in absorption mode FT-ICR MS is the baseline of the spectrum. Because of the rapid rate of phase accumulation, any error at the start of the transient is amplified, which can lead to baseline oscillation for high intensity peaks and negative intensity peaks for ions of low abundance, which makes them invisible for peak-picking algorithms. Baseline oscillation also occurs in 2D mass spectra, as can be seen in absorption mode narrowband 2D mass spectra.[18,19] One solution to this problem is to perform post-processing baseline correction. Another, less computational time-consuming solution, is the use of an asymmetrical apodization function proposed by Kilgour and Van Orden, which sets the initial datapoints of transients at zero.[32] Their proposed apodization function A has a rise and fall with a maximum at a fraction F of the transient length. Its equation is:

For $n \leq NF$:  $\qquad A(n) = \frac{1}{2}\left(1 - \cos\left(\frac{n\pi}{NF-1}\right)\right) \qquad$ (eq. 1)

For $n > NF$:  $\qquad A(n) = \frac{1}{2}\left(1 - \cos\left(\frac{(n+N(1-F)-1)\pi}{N(1-F)-1}\right)\right) \qquad$ (eq. 2)

in which $N$ is the number of datapoints in the transient, $F$ the fraction of transient at which the apodization function is at its maximum, $n$ the $n^{th}$ datapoint in the transient.

Here, we used a variant of the Kilgour-Van Orden apodization function. For $n > NF$, we used $\sqrt{|A(n)|}$ instead of the function in equation 2. We show this function in Fig. 2a with a maximum at $F = 0.25$, which we found to be optimal for our baseline correction and SNR.

Fig. 2b explains our choice of apodization function by showing the effect of apodization (red) on a randomly-chosen transient (black) from the 2D MS data set. The start of the transient is suppressed by the apodization to correct the baseline using equation 1. To limit the duration of 2D MS experiments, shorter transients are preferred during data acquisition. Hence, at the end of the transient, the decay of the signal is often cut short by the end of the acquisition. Therefore, we chose to apodize the transient with the square root of the asymmetric function in equation 2 to minimize signal loss at the transient end.

The result of the Fourier transformation in phase-corrected absorption mode (Fig. 2c, red) shows a corrected baseline compared to the same data processing with a standard shifted Sine Bell apodization (Fig. 2c, black). As a side note, the peaks, which represent the $[M+9H]^{9+}$ precursors and their isoforms, form abnormal isotopic distributions due to radius modulation (see Scheme 1). The peaks also show intense side-lobes that are similar to the ones presented by Kilgour and Van Orden, but which do not hinder data analysis.[32]

The signal in the vertical dimension also has a rapid rate of phase accumulation which cannot be corrected as easily as the transients. Instead, we chose to use an autoregressive model to correct errors in the initial datapoints. This model is based on the fact that, for a data set with $N$ datapoints, each real datapoint $x(n)$ at detection date or index $n$ can be expressed as the sum of the signal from the sample $\hat{x}(n)$ and the noise $e(n)$:

$$x(n) = \hat{x}(n) + e(n) \qquad \text{(eq. 3)}$$

In 2D MS, the signal $\hat{x}(n)$ can be expressed as a finite set of regularly-sampled, exponentially dampened sinusoids:[40]

$$\hat{x}(n) = \sum_{k=1}^{p} \alpha_k (z_k)^n \qquad \text{(eq. 4)}$$

where $\alpha_k$ are complex amplitudes (containing the phase information) and $z_k$ are the poles of the signal and $p$ is the number of poles:

$$z_k = e^{\gamma_k + j\omega_k} \qquad \text{(eq. 5)}$$

where $\omega_k$ are the frequencies of the signal and $\gamma_k$ the dampening factors. The signal can therefore be decomposed in p $[\alpha_k, z_k]$ vectors and is predictable. If $N > p$, then the value of $\hat{x}(n)$ can be predicted for $p < n \leq N$. Conversely, the datapoints between $N$-$p$ and $N$ can be used to calculate the datapoints for $1 \leq n \leq N - p - 1$.

In the vertical dimension of 2D mass spectra, $p$ corresponds approximately to the number of precursors a given fragment has. Therefore, typically, $p \leq 10$, for $N = 4096$. Here, the phase was found to rotate 392 times over the signal for the highest frequencies (see Supporting Information, configuration file for absorption mode data processing).[18] Since the phase in the vertical dimension accrues linearly, we estimated phase rotations started over the 392 first points of each vertical precursor ion scan and that they should be corrected. We plotted the result in Fig. 2d (red) for the precursor ion scan of the $c_{10}$ fragment and compared it to the precursor ion scan before baseline correction (black). As can be seen on Fig. 2d, especially for the peak corresponding to the $[M+10H]^{10+}$ precursor, the baseline correction somewhat suppresses the negative side-lobes, but, more importantly, suppresses the positive side-lobes which can be misinterpreted in the 2D mass spectrum. In addition, the SNR for $[M+10H]^{10+}$ increased from 15 to 20 with baseline correction, because the baseline fluctuation before correction was mistaken for noise.

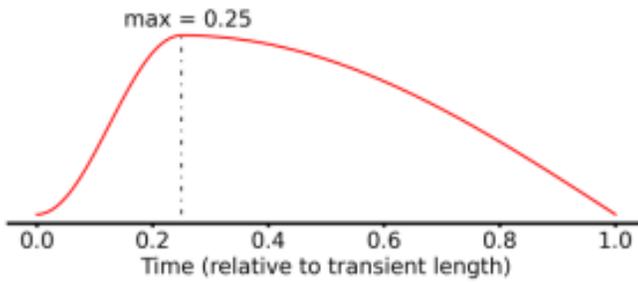

(a) Modified asymmetric apodization function

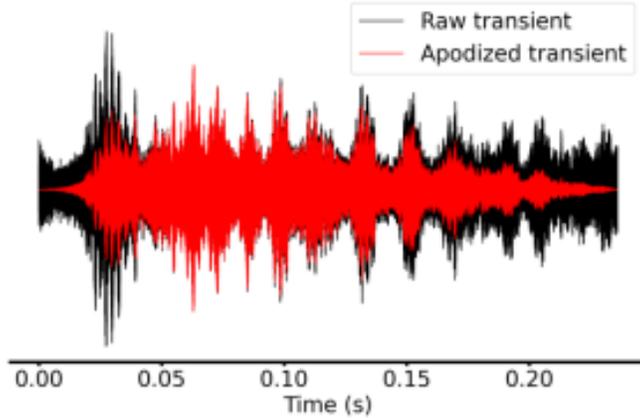

(b) Effect of apodization function on a transient

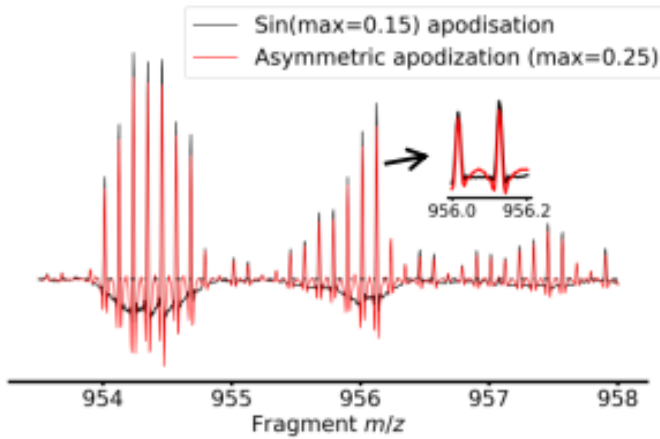

(c) Effect of apodization after Fourier transformation

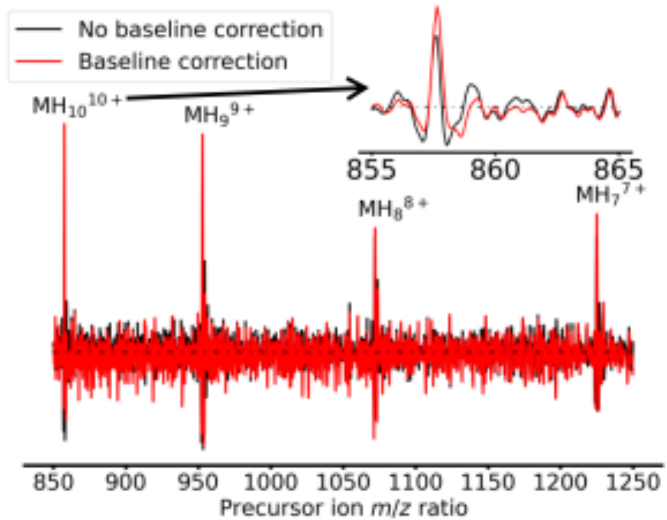

(d) Precursor ion scan of $c_{10}$ at $m/z$ 1136.65032

**Figure 2.** (a) Profile of the modified asymmetric apodization function proposed by Kilgour and Van Orden.[32] (b) Transient extracted from the 2D mass spectrum at $t_1$ = 1126 μs without apodization (black) and with the apodization from Fig. 2a (red). (c) Comparison between the absorption mode mass spectrum at $t_1$ = 1126 μs obtained with sinusoidal apodization (black) and the asymmetric apodization shown in Fig. 2a (red). (d) Precursor ion scan at *m/z* 1136.65032 without baseline correction (black) and with baseline correction (red). In (c-d) the dotted black line indicates zero intensity.

The improved resolving power and SNR in absorption mode 2D MS leads to higher confidence in peak assignment. In Fig. 3a, we show the fragmentation pattern of 10+ ubiquitin extracted from the 2D mass spectrum after data processing in magnitude mode (black) and in absorption mode (red). To show complete fragment ion isotopic distributions of $[M+10H]^{10+}$, we summed up 5 fragment ion scans between precursor ion *m/z* 857.3-857.7 to construct the spectra shown in Fig. 3a. Peak assignment was performed in FAST-MS without any modifications (all fragment assignments for these peaklists are reported in Tables S1 and S2 in the Supporting Information).[36] In Fig. 3a, peaks assigned in both magnitude and absorption mode are labelled in black, and those only assigned in absorption mode are labelled in red. Between fragment *m/z* 1100-1140 alone, 3 more fragments were assigned in absorption mode ($c_{20}^{2+}$, $z_{50}^{5+}$, and $z_{70}^{7+}$). After manual inspection of the automatic peak assignments, we found that the sequence coverage was 92% in absorption mode with 193 assigned fragments compared to 82% in magnitude mode with 146 assigned fragments (detailed peak assignment tables and sequence coverage in SI Tables S1 and S2 and Scheme S1 and S2). Mass accuracy reflects the confidence of peak assignments, and therefore we plotted normalized peak intensities vs. mass accuracy in Fig. 3c. In absorption mode (right), mass errors cluster closer to 0 ppm than in magnitude mode (left) with a smaller standard deviation (1.7 ppm vs. 2.7 ppm), thereby improving confidence in peak assignments. The absorption mode-induced increase of ion assignments observed for unmodified ubiquitin also applies to the fragments of oxidized ubiquitin, which thereby increases the structural resolution in mapping ubiquitin solvent accessibility.

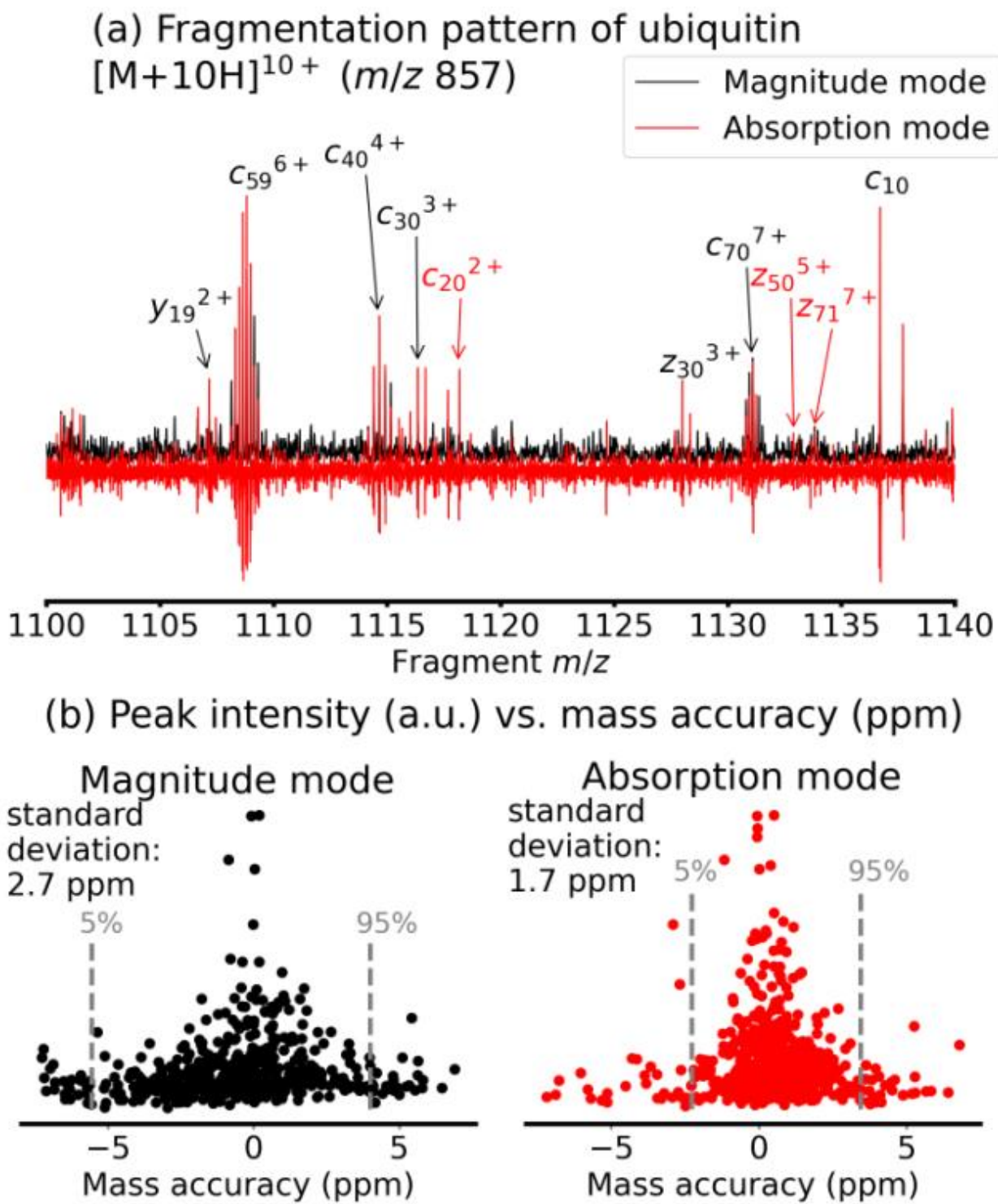

**Figure 3.** (a) Fragmentation pattern of ubiquitin [M+10H]$^{10+}$ obtained by the sum of fragment ion scans between precursor *m/z* 857.3-857.7 in magnitude mode (black) and absorption mode (red).

The peak assignments in black were found in both magnitude mode and absorption mode. The peak assignments in red were found only in absorption mode. (b) Normalized peak intensity vs. mass accuracy for assigned peaks in magnitude mode (left) and absorption mode (right).

We extracted fragment ion scans from the phase-corrected absorption mode 2D mass spectrum to construct the complete isotopic distribution of the fragmentation patterns of oxidized ubiquitin before peak-picking and peak assignment. Because the oxidation sites of ubiquitin in FPOP are *a priori* unknown, using externally calibrated peaklists for peak assignments can cause errors. For example, $y_n$ and $z_n$+ox fragments are isobaric and get more difficult to distinguish when their charge state increases. In 2D MS, the internal calibration for quadratic frequency-to-mass conversion of one fragment ion scan can be applied to all the others for high mass accuracy. Here, we applied the internal calibration of the 9+ charge state of unmodified ubiquitin to all the other fragment ion scans. FAST-MS performed peak assignments on the calibrated peaklists of ubiquitin with 1 or 2 oxidations at all 4 charge states that were detected in the 2D mass spectrum (10+-7+).[36] Oxidations were enabled at the following residues: F, H, I, K, L, M, P, R, V, W, Y.[23] As can be seen in the assignment examples shown in Fig. 4a, FAST-MS generated theoretical distributions that can be compared to peak *m/z* ratios and intensities for assignments.

As can be seen with the assignments of the isotopic distribution of $(c_{62}+ox)^{6+}$, $(z_{73}+ox)^{7+}$, and $c_{52}^{5+}$, FAST-MS is capble of making assignments of overlapping isotopic distributions. Because we chose to sum up 5 fragment ion scans for each precursor ion, we were not able to recover the full isotopic distributions of each fragment ion, hence the discrepancies observed between the theoretical and the experimental isotopic distributions in fig. 4a.

A one-dimensional mass spectrum of the oxidized ubiquitin (see Fig. S2) was deconvolved in the DataAnalysis 6.0 software (Bruker Daltonics, Bremen, Germany) using the Sophisticated

Numerical Annotation Procedure (SNAP) algorithm.[41] The overall extent of oxidation was determined to be 40% for one oxidation and 28% for two oxidations. As a result, fragment ion abundances and therefore SNR are lower for oxidized ubiquitin species, since they are dependent on precursor ion abundances, i.e. amount of incorporated oxygen. For ubiquitin with one oxidation, the sequence coverage was found to be between 28 and 50% according to charge state, as shown in Fig. 4b. Similarly, for ubiquitin with 2 oxidations, the sequence coverage was found to be between 15 and 28% (peaklists are available in the Supporting Information).

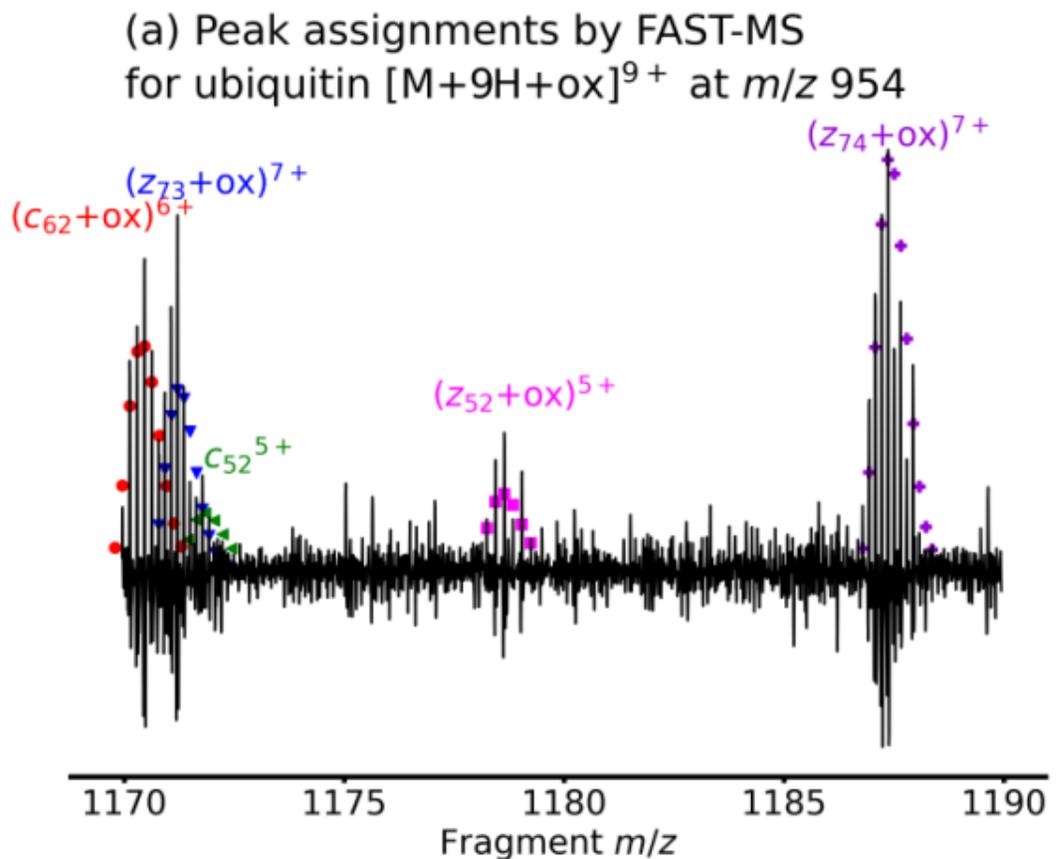

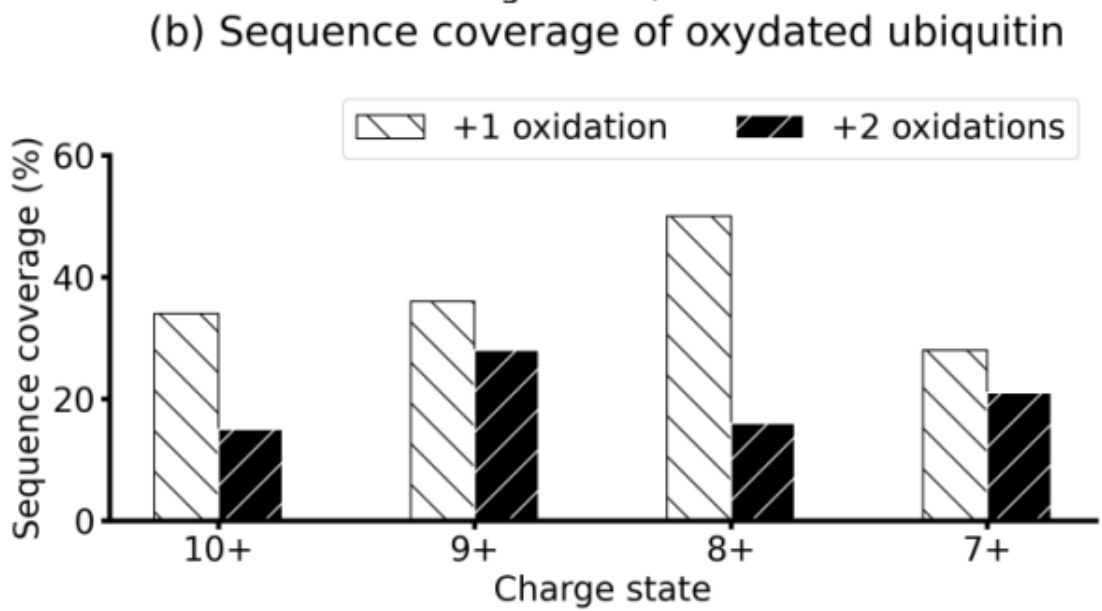

**Figure 4.** Data analysis of the 2D mass spectrum in phase-corrected absorption mode. (a) Extracted fragment ion scan of $[M+9H+ox]^{9+}$ at *m/z* 954 and peak assignments by FAST-MS (colored dots). (b) Sequence coverage of oxidized ubiquitin after peak assignment by FAST-MS

for the extracted fragment ion scans after reconstruction of the isotopic distribution for ubiquitin with 1 and 2 oxidations for all charge states in the 2D mass spectrum.

Top-down proteomics is not the only application in which analytes with chemical similarities yield fragment ions with the same *m/z*. Mass spectra of samples in metabolomics and the analysis of natural compounds tend to also be complex with many peaks close in mass and chemical composition, which can produce precursor ion scans with multiple peaks in 2D MS. This property makes them eminently suitable for broadband absorption mode 2D MS. A convenient example for testing our algorithm is the 2D UVPD mass spectrum of an extract of ergot alkaloids, shown in Figure 5. Ergot alkaloids are found as fungal growth-induced contamination in cereal products and may have toxic (e.g. ergotism) or pharmacologically beneficial effects upon ingestion.[42] They are composed of an ergoline ring that is methylated on the N-6 nitrogen and substituted with a peptidic ring on C-8.[43]

To reduce the intensity of the ergocristine peak ($C_{35}H_{40}N_5O_5^+$) at *m/z* 610.302396, which would otherwise have dominated the mass spectrum, CASI isolation was applied in the front-end of the instrument for controlled peak suppression.[29] The mass spectrum of the ergot alkaloid extract (shown in full in Figure S6 in the Supporting Information) enabled the assignment of multiple structurally similar ergot alkaloids in the sample, such as ergocryptine, ergotamine, ergocristine, and ergocristine dehydrate (see peak assignments in Table S12 of the Supporting Information).[44] All ions in the mass spectrum and 2D mass spectrum were singly-charged. In addition to these known compounds, we assigned a novel peak at *m/z* 592.255421 to $C_{34}H_{34}N_5O_5^+$. This compound is probably an impurity.

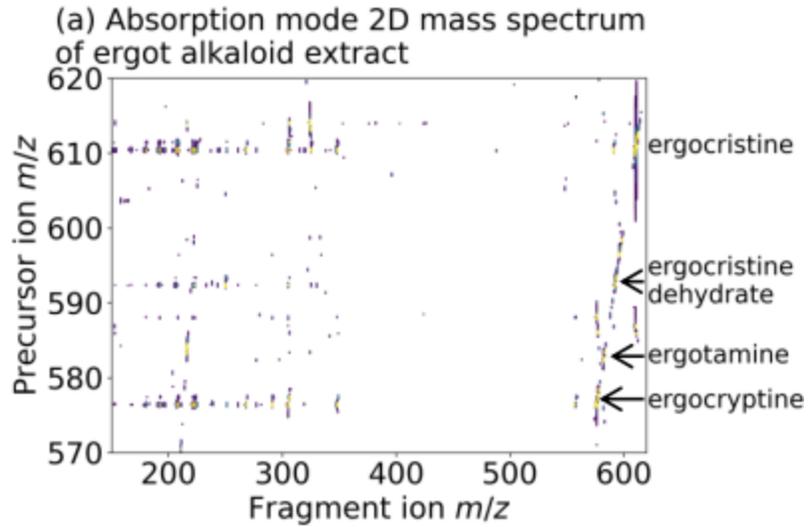
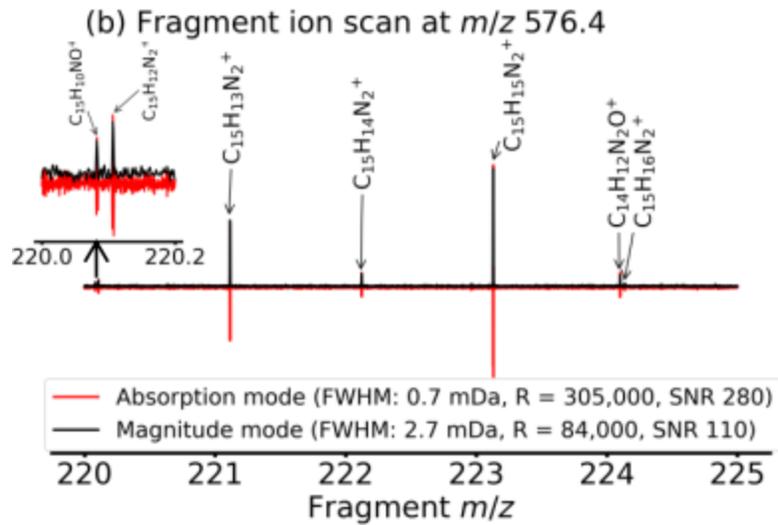
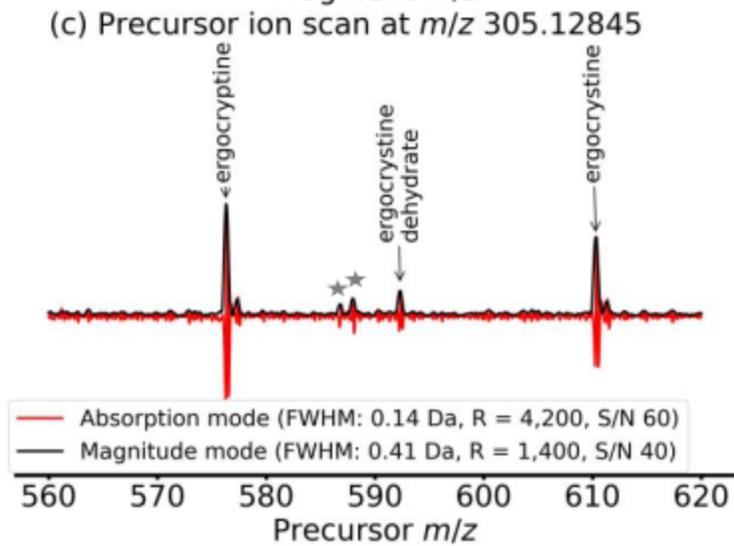

**Figure 5.** (a) Phase-corrected absorption mode 2D UVPD mass spectrum of an extract of ergot alkaloids (with SANE denoising). (b) Extracted fragment ion scan of ergocryptine ($C_{35}H_{40}N_5O_5^+$ at *m/z* 576.318046): comparison between absorption mode and magnitude mode (no denoising). (c) Extracted precursor ion scan of $C_{19}H_{17}N_2O_2^+$ (*m/z* 305.128454): comparison between absorption mode and magnitude mode (no denoising).

From the same absorption mode-processed 2D mass spectrum shown in Fig. 5a, we were able to extract the fragmentation patterns of ergocristine, ergocristine dehydrate, and ergocryptine, which correspond to the results found in previous studies by Lehner *et al.*[44] We used a precursor ion scan with multiple known peaks to calculate the coefficients of the linear phase correction function. Fig. 5b shows the comparison between the fragment ion scan of ergocryptine extracted at *m/z* 576.4 in absorption mode and in magnitude mode. Measurements of the resolving power and the SNR were taken for the peak at *m/z* 223.123036. The magnitude mode 2D mass spectrum was processed with one zerofill horizontally compared to two zerofills in absorption mode, which explains the 3.6 factor improvement in resolving power instead of the factor of approximately 2 that is expected. The average mass error of peak assignments also decreased from an 0.35 ppm in magnitude mode to 0.10 ppm in absorption mode (see peak assignments in Tables S13 to S20 in the Supporting Information). In the precursor ion dimension, the resolving power and SNR both increase from magnitude mode to absorption mode (see Fig. 5c) by a factor of 3 and 1.5 respectively.

Due to baseline isotopic resolution in the vertical precursor ion dimension, we were able to perform peak-picking and centroiding in two dimensions. Doing so allowed us to measure both precursor and the fragment *m/z* of all fragment peaks. For the fragments of ergocryptine (*m/z* 576.318037), the difference between maximum and minimum precursor *m/z* was 220 mDa in magnitude mode (standard deviation: 38 mDa) and 45 mDa in absorption mode (standard deviation: 6 mDa). For the fragments of ergocristine (*m/z* 610.302337), the range of precursor *m/z* measurements was 111 mDa in magnitude mode (standard deviation: 22 mDa) and 24 mDa in absorption mode (standard deviation: 3 mDa). These results are consistent with precursor *m/z* measurements reported by Marzullo *et al.*[8]

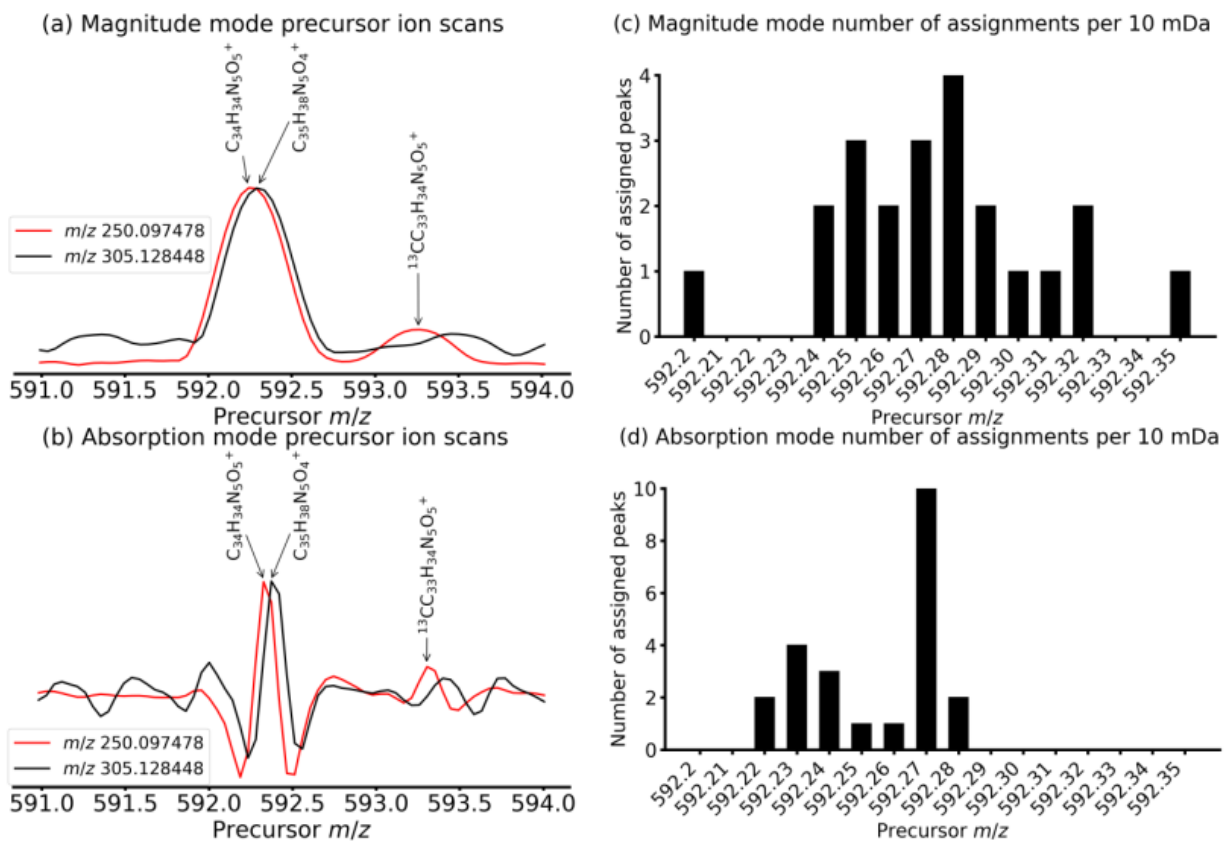

**Figure 6.** Comparison between the normalized precursor ion scans of the fragment at *m/z* 250.097478 ($C_{15}H_{12}N_3O^+$) and the fragment at *m/z* 305.128448 ($C_{19}H_{17}N_2O_2^+$) extracted from the (a) magnitude mode and (b) absorption mode 2D mass spectrum of the ergot alkaloid extract. Number of peak assignments per 0.01 Da for peaks detected in the precursor *m/z* 592-593 and fragment *m/z* 120-600 region of the (c) magnitude mode and (d) absorption mode 2D mass spectrum of the ergot alkaloid extract.

When checking against the autocorrelation line and the one-dimensional mass spectrum, the fragments at precursor *m/z* 592.3 have two possible precursors: $C_{35}H_{38}N_5O_4^+$ at *m/z* 592.291815 (ergocristine dehydrate) and $C_{34}H_{34}N_5O_5^+$ at *m/z* 592.255421 (unknown), which have a mass difference of 36 mDa ($CH_4$ vs. O), which would not be separable with standard quadrupolar precursor isolation. In Fig. 6a, we plotted the normalized precursor ion scans of *m/z* 250.097478 (assigned to $C_{15}H_{12}N_3O^+$) and *m/z* 305.128448 (assigned to $C_{19}H_{17}N_2O_2^+$). The former fragment only has one precursor at *m/z* 592.3 and the latter fragment has precursors at *m/z* 576.3, *m/z* 592.3, and *m/z* 610.3 (see Fig. 5c). The noise level for the precursor ion scan of *m/z* 305.128448, resulting mainly from scintillation noise which is proportional to the intensity of the signal, is therefore higher than the noise level of the precursor ion scan of *m/z* 250.097478.[40] In Fig. 6a, the two peaks are shifted, suggesting that the precursor for *m/z* 250.097478 is $C_{34}H_{34}N_5O_5^+$ and the precursor for *m/z* 305.128448 is $C_{35}H_{38}N_5O_4^+$ (ergocristine dehydrate). However, the vertical resolving power in magnitude mode (R = 1,300 at *m/z* 592.3) is not sufficient to confidently assign precursor ions. In contrast, after absorption mode processing, the vertical resolving power dramatically increases to 6,200 at *m/z* 592.3, as plotted in Fig. 6b. Here, the shift between the two peaks is more pronounced and we can therefore confirm the assignments with increased confidence.

To further distinguish between the two precursor ions, we sorted the assigned fragment ion peaks in the precursor *m/z* 592.20-592.35 range in *m/z* 0.01 bins. In Fig. 6c, the precursor *m/z* of the fragment ion peaks from the magnitude mode 2D mass spectrum are within an *m/z* 592.20-592.35 range and the two populations overlap. We therefore cannot distinguish between the fragments of $C_{35}H_{38}N_5O_4^+$ and $C_{34}H_{34}N_5O_5^+$. In Fig. 6d, however, the precursor *m/z* of the fragment ion peaks from the absorption mode 2D mass spectrum are within an *m/z* 592.22-592.29 range. Two separate ion populations can be distinctly observed, and the fragments of $C_{35}H_{38}N_5O_4^+$ and $C_{34}H_{34}N_5O_5^+$ can therefore be clearly distinguished. The average m/z of the two ion populations is shifted by approximately m/z 0.02 due to an inaccuracy in the offset of the modulation frequency in the vertical dimension reported in the Bruker software.[1]

At 15 T magnetic field, the difference in cyclotron (and modulation) frequency between the two precursor ions is 24 Hz. The distance between points in the vertical precursor ion dimension in both the absorption and magnitude mode 2D frequency spectrum is 31 Hz. A previous study showed that accurate correlation between precursor and fragment ions was possible for a difference in precursor cyclotron frequency that was less than the distance between points in the vertical dimension.[19] We can see in both Figs. 6a and 6b that the two precursor ion scans are shifted by one datapoint, which is consistent with the cyclotron frequency difference between the two precursor ions. However, the superior resolving power in absorption mode probably reduces the error in measuring peak centroids compared to magnitude mode, which in turn improves the confidence with which we can correlate precursor and fragment ions, clearly demonstrating the benefits of broadband absorption mode 2D MS for complex samples with many peaks closely spaced in *m/z*.

CONCLUSION

Removing the limitation of data size, imposed by computational memory, by adopting batch-processing, enabled us to successfully phase-correct a broadband 2D mass spectrum of FPOP-oxidized ubiquitin in both dimensions for absorption mode, as a model top-down MS study of a protein with many proteoforms. We were also able to correct the baseline of the 2D mass spectrum in both the precursor and fragment ion dimension, with a variant on the Kilgour-Van Orden asymmetrical apodization, and in the precursor ion dimension, with an autoregressive denoising algorithm. When comparing the resulting absorption mode 2D mass spectrum with the magnitude mode 2D mass spectrum processed from the same data set, we demonstrate improved SNR, resolving power, mass accuracy and ultimately peak-assignment and protein sequence coverage.

We also applied our approach and phase-corrected absorption mode batch-processing to the 2D UVPD mass spectrum of an extract of ergot alkaloids. Due to their structural similarities, the compounds in the sample have common fragments whose signals can be used to calculate phase correction coefficients in the precursor ion dimension. Here we also observed increased resolving powers and SNR by a factor of 2-3. The increased resolving power in particular enabled the accurate correlation between precursor and fragment ions for two precursor ions with *m/z* differences as low as 36 mDa.

The optimization of the coefficients of the phase correction functions remains onerous and would gain from automation.[16] In a future study, we will also investigate the re-use of processing parameters for multiple 2D mass spectra acquired in identical experimental conditions.[14] However, already the two presented examples of very diverse applications clearly demonstrate how absorption mode data processing improves the performance of 2D MS, and the strengths that FT-ICR phasing and absorption mode processing offer for highly complex samples.

ASSOCIATED CONTENT

Supporting information, containing peaklists with assignments and additional figures (PDF file)

Phase-corrected absorption mode data processing program (python program file)

Data processing parameter files (text files)

Raw data files (miscellaneous files)

AUTHOR INFORMATION

Corresponding Author

Maria A. van Agthoven : maria.van-agthoven@univ-rouen.fr

Present Addresses

Marek Polák: ICR Program, National High Magnetic Fields Laboratory, 1800 East Paul Dirac Drive, Tallahassee FL32310, United States

Jan Fiala: Thermo Fisher Scientific, Achtseweg Noord 5, 5651 GG Eindhoven, Netherlands

Michael Palasser: Bachem AG, Hauptstrasse 144, 4416 Bubendorf, Switzerland

Author Contributions

The manuscript was written through contributions of all authors. All authors have given approval to the final version of the manuscript.

Funding Sources

This project has received funding from the European Union's Horizon 2020 research and innovation programme under the Marie Skłodowska-Curie grant agreement No 101034329. This project has received funding from the European Union's Horizon 2020 research and innovation programme under grant agreement No 731077. M.v.A. and K.B. thank the Austrian Science Fund for the Lise Meitner Fellowship M2757-B. This work was also financially supported by the Czech Science Foundation (25-18181S), the National Institute for Neurological Research (Programme EXCELES, ID Project No. LX22NPO5107), the Ministry of Education, Youth and Sports of the Czech Republic grant PHOTOMACHINES- Photosynthetic cell redesign for high yields of therapeutic peptides (CZ.02.01.01/00/22_008/ 0004624) ) and INTER-MICRO - Talking with microbes – understanding microbial interactions within One Health framework (CZ.02.01.01/00/22_008/0004597) and the Academy of Sciences of the Czech Republic (RVO: 61388971) dedicated to P.N.

Notes

Any additional relevant notes should be placed here.

ACKNOWLEDGMENT

M.v.A. dedicates this article to the memory of Ms. Elly Derks. The authors thank Dr. Miroslav Flieger for providing the ergot alkaloid sample. This project has received funding from the European Union's Horizon 2020 research and innovation programme under the Marie Skłodowska-Curie grant agreement No 101034329. This project has received funding from the European Union's Horizon 2020 research and innovation programme under grant agreement No 731077. M.v.A. and K.B. thank the Austrian Science Fund for the Lise Meitner Fellowship M2757-B. This work was also financially supported by the Czech Science Foundation (25-18181S), the National Institute for Neurological Research (Programme EXCELES, ID Project No. LX22NPO5107), the Ministry of Education, Youth and Sports of the Czech Republic grant PHOTOMACHINES- Photosynthetic cell redesign for high yields of therapeutic peptides (CZ.02.01.01/00/22_008/ 0004624) and INTER-MICRO - Talking with microbes – understanding microbial interactions within One Health framework (CZ.02.01.01/00/22_008/0004597) and the Academy of Sciences of the Czech Republic (RVO: 61388971) dedicated to P.N.

ABBREVIATIONS

2D MS, two-dimensional mass spectrometry;FPOP, fast photochemical oxidation of proteins; FT-ICR MS, Fourier transform ion cyclotron resonance mass spectrometry; ECD, electron capture dissociation; UVPD, ultraviolet photodissociation; SNR, signal-to-noise ratio.

Graphical abstract:

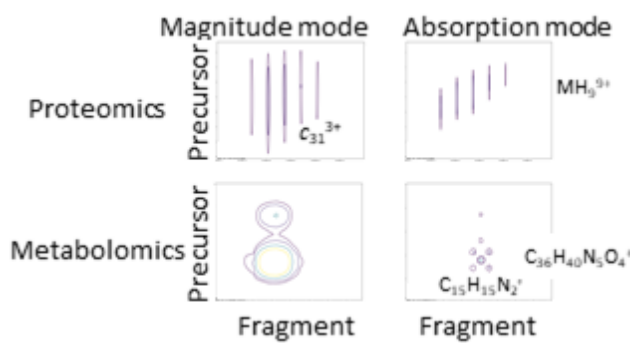